\documentclass[aps,pra,showpacs,amsmath,amssymb,amsfonts,superscriptaddress,twocolumn,lengthcheck]{revtex4-2}
\usepackage [latin1]{inputenc}
\usepackage{amsmath}
\usepackage{amsfonts}
\usepackage{graphicx}
\usepackage{color}
\usepackage[colorlinks,citecolor=blue]{hyperref}
\usepackage{titlesec}
\titleformat{\section}{\small\bfseries\centering\MakeUppercase}{\thesection}{1em}{}
\begin{document}

\title{Chiral-cat-state generation via the Sagnac-Fizeau effect}
\author{Yu-Hong Liu}
\affiliation{Key Laboratory of Low-Dimensional Quantum Structures and Quantum Control of Ministry of Education, Key Laboratory for Matter Microstructure and Function of Hunan Province, Department of Physics and Synergetic Innovation Center for Quantum Effects and Applications, Hunan Normal University, Changsha, 410081, China}
\author{Xian-Li Yin}
\affiliation{Key Laboratory of Low-Dimensional Quantum Structures and Quantum Control of Ministry of Education, Key Laboratory for Matter Microstructure and Function of Hunan Province, Department of Physics and Synergetic Innovation Center for Quantum Effects and Applications, Hunan Normal University, Changsha, 410081, China}
\author{Hui Jing}
\affiliation{Key Laboratory of Low-Dimensional Quantum Structures and Quantum Control of Ministry of Education, Key Laboratory for Matter Microstructure and Function of Hunan Province, Department of Physics and Synergetic Innovation Center for Quantum Effects and Applications, Hunan Normal University, Changsha, 410081, China}
\author{Le-Man Kuang}
\affiliation{Key Laboratory of Low-Dimensional Quantum Structures and Quantum Control of Ministry of Education, Key Laboratory for Matter Microstructure and Function of Hunan Province, Department of Physics and Synergetic Innovation Center for Quantum Effects and Applications, Hunan Normal University, Changsha, 410081, China}
\author{Jie-Qiao Liao}
\email{Contact author: jqliao@hunnu.edu.cn}
\affiliation{Key Laboratory of Low-Dimensional Quantum Structures and Quantum Control of Ministry of Education, Key Laboratory for Matter Microstructure and Function of Hunan Province, Department of Physics and Synergetic Innovation Center for Quantum Effects and Applications, Hunan Normal University, Changsha, 410081, China}
\affiliation{Institute of Interdisciplinary Studies, Hunan Normal
University, Changsha, 410081, China}

\begin{abstract}
Chiral quantum state generation is an interesting topic in quantum physics and quantum information science. Here we propose an approach for generating chiral cat states in a spinning resonator supporting both the clockwise (CW) and counterclockwise (CCW) traveling modes, which are dispersively coupled to a two-level atom. The physical mechanism for the chiral-cat-state generation is based on the  Sagnac-Fizeau effect. Concretely, when the resonator is rotating, the CW and CCW modes have different frequency detuning with respect to the atomic transition frequency and hence the atomic-state-dependent rotating angular velocities for the CW and CCW modes in phase space are different. This mode-dependent evolution leads to a chirality mechanism in the state generation. Based on the mode-dependent conditional rotation evolution  and  atomic projection  measurement at proper time, we achieve the separate generation of cat states in the CW mode and coherent states in the CCW mode. We also investigate quantum coherence properties of the generated states by examining their Wigner functions. In addition, the influence of the system dissipations on the state generation in the open-system case is investigated. Our work will provide some insights into the development of chiral optical devices and nonreciprocal photonics.
\end{abstract}

\date{\today}
\maketitle

\section{INTRODUCTION}
The Schr\"{o}dinger cat states are an essential resource for studying fundamental quantum
physics~\cite{Schroedinger1935,Peres2002}, including exploring macroscopic quantum coherence~\cite{LeggettPRL1985,LeggettJCM2002}, quantum decoherence~\cite{ZurekRMP2003,Joos2003,BrunePRL1996}, and quantum-classical boundary~\cite{ZurekPT1991,HarocheRMP2013}.~They also have wide application potential in quantum information processing, such as quantum teleportation~\cite{LiaoPLA2006,FurusawaSci1998,BraunsteinPRL1998}, quantum communication~\cite{JeongPRA2001,BraunsteinRMP2005,VlastakisSc2013,Gilchrist2004}, quantum computation~\cite%
{CochranePRA1999,RalphPRA2003,MirrahimiNJP2014,Ofekature2016,Guillaud2023,JeongPRA2002}, and quantum precision
measurement~\cite{JooPRL2011,MunroPRA2002,BildSci2023}. So far, various methods have been proposed
to create the Schr\"{o}dinger cat states, including conditional rotation mechanism~\cite{VlastakisSc2013,GuoPLA1996,BrunePRA1992}, conditional
displacement mechanism~\cite%
{MonroeSciece1996,BosePRA1997,ArmourPRL2002,AshhabPRA2010,LiaoPRL2016,LiaoPRA2016,ChenPRL2021}%
, non-Gaussian measurement~\cite{OurjoumtsevNa2007,SunPRL2021,HuPRR2023,OurjoumtsevScien2006,TakahashiPRL2008,HanLPR2023}, quantum state
transfer~\cite{HoffPRL2016,TehPRA2018}, Kerr-nonlinear evolution~\cite{MiranowiczQO1990,KuangPRA2003,LiaoPRA2020,GrimmNa2020,YinPRA2022,HeNC2023}, and reservoir engineering
techniques~\cite{PoyatosPRL1996,AsjadASjas2014,BrunelliPRA2018}.~Motivated by the rapid development of chiral and nonreciprocal quantum devices, it becomes an interesting topic to prepare chiral quantum resources~\cite{LodahlA2017,Hafezi2024}, including chiral quantum superposition, entanglement, and squeezing. These resources are important elements for implementing nonreciprocal application such as nonreciprocal quantum communications~\cite{FengSC2011,ScheucherSC2016}, chiral
optical devices~\cite{SounasNP2017,TangPRL2022,ZuoOE2024}, and stable codes for quantum error correction~\cite{TerhalRMP2015,GeorgescuNEP2020}. Consequently, how to create chiral cat states, which exhibit mirror asymmetry in quantum  superposition, remains an interesting and desired task in chiral quantum optics~\cite{LodahlA2017,Hafezi2024} and nonreciprocal photonics~\cite{SounasNP2017}. Note that the chirality, which is based on the spatial reflection asymmetry~\cite{Sakurai1994}, has been suggested to implement nonreciprocal quantum devices~\cite{LiSC2024,YouPRA2021}, in which the nonreciprocity is usually introduced by the breaking of the time-reversal symmetry~\cite{Sakurai1994}.

Quantum nonreciprocal devices are fundamental components in chiral quantum
technologies and much recent attention has been paid to nonreciprocal physics~\cite%
{YangPRL2019,JiaoPRL2020,DongSC2021,XuPRAP2020,XuPRA2021,MetelmannPRX2015,FruchartNa2021}. The spinning resonators~\cite{MaayaniNa2018,MalykinPU2000,JingOpti2018}, as a typical nonreciprocal device~\cite
{YangPRL2019,JiaoPRL2020,DongSC2021,XuPRAP2020,XuPRA2021,MetelmannPRX2015,FruchartNa2021,MaayaniNa2018,MalykinPU2000,JingOpti2018}, could be used to provide an ideal platform for studying chiral quantum phenomena due to the equivalence between  the spatial  reflection and the time-reversal operation in this system. Recently, significant
progress has been achieved based on spinning resonators, including nonreciprocal photon blockade~\cite%
{HuangPRL2018}, nonreciprocal entanglement~\cite{JiaoPRL2020,JiaoPRAA2022}, and nonreciprocal topological phonon transfer~\cite{LaiPRL2024}. However, the controlled generation and manipulation of chiral quantum states remain to be explored. Addressing these issues is crucial for both understanding and practical applications of chiral quantum technologies.

In this paper, we explore the generation of chiral cat states in a bichromatic Jaynes-Cummings system~\cite{ZhuPRL2024,JingPRA2021,WangPRA2017} comprising a two-level atom coupled to a spinning resonator, which supports both clockwise (CW) and counterclockwise (CCW) traveling modes. The rotation of the resonator induces a Sagnac-Fizeau effect~\cite{MalykinPU2000}, resulting in different effective detunings for the CW and CCW modes relative to the atomic frequency. We consider the large detuning case between the atom and the two cavity modes, which enables a dispersive interaction. The different frequency detunings of the CW and CCW modes will induce different conditional rotation in phase space for the two traveling modes, which can be used to provide a chirality mechanism for generating distinct states. Concretely, by performing a projection measurement on the atom at a specific time, a cat state can be prepared in the CW mode and a coherent state in the CCW mode, or different cat states can be generated in both modes at other measurement times. Here, the simultaneous preparation of different quantum states in the CW and CCW modes of the spinning resonator is referred to as chiral quantum state generation. We further investigate the coherence properties of these generated states by examining their Wigner functions. We also consider the influence of the system dissipations on the state generation. Our proposal paves the way for developing quantum chiral devices based on the coupled atom-cavity systems.

\begin{figure}[bp]
\centering \includegraphics[width=0.44\textwidth]{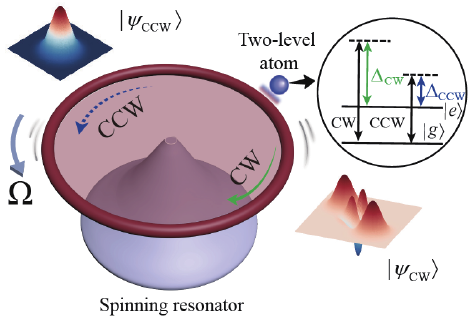}
\caption{Schematic of the cavity-QED system consisting of a two-level atom coupled to a spinning
resonator via the optical evanescent fields. The spinning resonator supports both the clockwise (CW) and
counterclockwise (CCW) traveling modes, denoted as $a_{\mathrm{CW}}$ and $a_{%
\mathrm{CCW}}$, respectively. The two cavity modes are initially prepared in coherent states. By rotating the resonator along the CCW direction at an angular velocity $\Omega$, the effective frequency for the CW (CCW)
traveling mode experienced by the atom becomes $\protect\omega_{c}+\Delta_{\mathrm{sag}}$ ($\protect%
\omega_{c}-\Delta_{\mathrm{sag}}$), where $\protect\omega_{c}$ is the
nonspinning resonance frequency of the resonator and $\Delta_{\mathrm{sag}}$
is the rotation-induced Sagnac-Fizeau frequency shift.}
\label{Fig1}
\end{figure}
The rest of this paper is organized as follows. In Sec.~\ref{Sec2}, we
introduce the physical model and derive the effective Hamiltonians. In Sec.~\ref{sec3}, we investigate the generation
of chiral cat states and analyze the nonclassical properties of these generated states by examining their Wigner functions. We also
numerically calculate the exact state generation and verify the
validity of the approximation with the fidelity. In Sec.~\ref{sec4}, we present
numerical simulations of the state generation in the open-system case and
investigate the influence of the system dissipations on the state generation. Finally, we
discuss the experimental implementation of this scheme and conclude this
work in Sec.~\ref{sec5}.

\section{Physical model and Hamiltonians\label{Sec2}}
We consider a cavity-QED system consisting of a two-level atom with an energy separation $%
\omega_{a}$ coupled to a spinning resonator supporting both the CW and CCW traveling
modes via the optical evanescent fields (Fig.~\ref{Fig1}). The rotation of the resonator in the CCW direction at an angular velocity $\Omega$ induces a Sagnac-Fizeau frequency shift for both the CW and
CCW modes. Then the effective frequencies of the CW and CCW modes become $\omega_{c}+\Delta_{\mathrm{sag}}$ and $\omega_{c}-\Delta_{\mathrm{sag}}$, respectively. Here, $\omega_{c}$ is
the resonance frequency of the stationary resonator and $\Delta_{\mathrm{sag%
}}$ is the Sagnac-Fizeau frequency shift defined as~\cite{MalykinPU2000}
\begin{equation}  \label{eq1}
\Delta_{\mathrm{sag}}=\frac{n_{r}R\Omega\omega_{c}}{c}\left(1-\frac{1}{%
n_{r}^{2}}-\frac{\lambda}{n_{r}}\frac{dn_{r}}{d\lambda}\right).
\end{equation}
In Eq.~(\ref{eq1}), $\lambda$ and $c$ represent the wavelength of the field mode and the speed of light in vacuum, respectively, $n_{r}$ is the refractive index, and $R$ is the radius of the resonator. The term $dn_{r}/d\lambda$ is the dispersion term related to the relativistic origin of the Sagnac effect, which is relatively small and can be ignored in typical materials~\cite{MaayaniNa2018,MalykinPU2000}.

In a rotating frame defined by the unitary operator $\mathrm{exp}%
[-i\omega_{a}(\sigma_{z}/2+a_{\mathrm{CW}}^{\dagger}a_{\mathrm{CW}%
}+a_{\mathrm{CCW}}^{\dagger}a_{\mathrm{CCW}})t]$, the
Hamiltonian of the system can be written as ($\hbar=1$)
\begin{eqnarray}  \label{eq2}
H=\sum_{\epsilon=\mathrm{CW},\mathrm{CCW}}[\!\Delta_{\epsilon}a_{\epsilon}^{%
\dagger}a_{\epsilon}+J(a_{\mathrm{\epsilon}%
}^{\dagger}\sigma_{-}+a_{\mathrm{\epsilon}}\sigma_{+})],
\end{eqnarray}
where $\sigma_{z}\equiv |e\rangle\langle e|-|g\rangle\langle g|$ is $z$-direction Pauli
operator, with $|g\rangle$ and $|e\rangle$ being the ground and
excited states of the two-level atom, respectively. The operators $\sigma_{-}\equiv|g\rangle\langle e|$ and $\sigma_{+}\equiv|e\rangle\langle g|$ are, respectively, the lowering and raising operators of the atom. The operators $a_{\epsilon}^{\dagger}$ and $%
a_{\epsilon}$ ($\epsilon=\mathrm{CW},\mathrm{CCW}$) are the creation and
annihilation operators of the $\epsilon$ mode. The variables $\Delta_{%
\mathrm{CW}}\equiv \Delta+\Delta_{\mathrm{sag}}$ and $\Delta_{%
\mathrm{CCW}}\equiv \Delta-\Delta_{\mathrm{sag}}$ represent, respectively, the effective frequency detunings of the CW and CCW modes with respect to the resonance frequency of the atom, with  $\Delta\equiv\omega_{c}-\omega_{a}$. The parameter $J$ denotes the coupling strength between the two traveling modes and the atom. We note that the bichromatic Jaynes-Cummings described by the Hamiltonian~(\ref{eq2}) can be realized in various physical systems such as coupled whispering-gallery microresonators-atom systems~\cite{AltonNP2011,JungePRL2013,ScheucherSC2016} and circuit QED systems~\cite{StrauchPRL2010,MaSL2014}. For the purpose of chiral state generation, we consider the coupled atom-spinning-resonator system in this work.

In this work, we consider the dispersive regime of the cavity-field-atom couplings, namely, the detunings between the two-level atom and the cavity field are much larger than the coupling strengths, i.e., $\Delta_{\epsilon=\mathrm{CW},\mathrm{CCW}}\gg J\sqrt{n_{\epsilon}+1}$~\cite{Schleich2001}, where $n_{\epsilon}$ are the maximally involved photon numbers in the two cavity modes. In this case, we can perform the Frohlich-Nakajima
transformation~\cite{FPR1950,Nakajima1955} on the Hamiltonian~(\ref{eq2}) to derive the effective Hamiltonian by eliminating the coherent transitions between the two atomic states,
\begin{eqnarray}  \label{eq4}
H_{S} \!\!&=\!\!&e^{-S}He^{S}  \notag \\
\!\!&\approx\!\! &\sum_{\epsilon =\mathrm{CW},\mathrm{CCW}}(\Delta _{\epsilon
}a_{\epsilon }^{\dagger }a_{\epsilon }\!-\!\zeta_{\epsilon}a_{\epsilon }^{\dagger }a_{\epsilon }\sigma _{z}\!-\!\zeta_{\epsilon}\sigma _{z}/2)  \notag \\
&&\!-\!\frac{1}{2}(\zeta_{\mathrm{CW}}\!+\!\zeta_{\mathrm{CCW}})
(a_{\mathrm{CW}}^{\dagger }a_{\mathrm{CCW}}\!+\!a_{\mathrm{CCW}%
}^{\dagger }a_{\mathrm{CW}})\sigma _{z},
\end{eqnarray}
where $S=\sum_{\epsilon=\mathrm{CW},\mathrm{CCW}}(J/\Delta_{\epsilon})(a_{\mathrm{%
\epsilon}}\sigma_{+}-a_{\mathrm{\epsilon}}^{\dagger}\sigma_{-})$
with $\zeta_{\epsilon}=J^{2}/\Delta_{\epsilon}$.
Under the parameter condition
\begin{equation}  \label{eq5}
\zeta_{\epsilon =\mathrm{CW},\mathrm{CCW}}/(4\Delta_{\mathrm{%
sag}})\ll1,
\end{equation}
we can discard approximately the excitation-exchange interaction term between the two
traveling modes and obtain the approximate Hamiltonian
\begin{equation}  \label{eq6}
H_{\mathrm{app}}= \sum_{\epsilon =\mathrm{CW},\mathrm{CCW}} (\Delta
_{\epsilon }a_{\epsilon }^{\dagger }a_{\epsilon }-\zeta_{\epsilon}a_{\epsilon }^{\dagger }a_{\epsilon }\sigma _{z}-\zeta_{\epsilon}\sigma _{z}/2).
\end{equation}

The physical mechanism of the chiral-cat-state generation can be understood more
clearly by expressing the approximate Hamiltonian of the system in the form
of conditional rotation couplings
\begin{eqnarray}
H_{\mathrm{app}} &=&\sum_{\epsilon =\mathrm{CW},\mathrm{CCW}}[\Delta
_{\epsilon }a_{\epsilon }^{\dagger }a_{\epsilon }-\zeta_{\epsilon}( a_{\epsilon }^{\dagger }a_{\epsilon }+1/2) \vert e\rangle
\langle e\vert  \notag \\
&&+\zeta_{\epsilon}( a_{\epsilon
}^{\dagger }a_{\epsilon }+1/2
) \vert g\rangle \langle g\vert ].
\end{eqnarray}
Here, the opposite signs of the coefficients for $|e\rangle\langle e|$ and
$|g\rangle\langle g|$ can be used to generate cat states for the traveling modes and
the different values of $\zeta_{\mathrm{CW}}$ and $\zeta_{\mathrm{CCW}}$ lead to different phase-space positions for the two traveling modes, since the effective rotating velocities of the two modes are different.

\section{Generation of the chiral cat states\label{sec3}}

In this section, we explore the generation of chiral cat states based on the approximate Hamiltonian $H_{\mathrm{app}}$ in
Eq.~(\ref{eq6}). To investigate the quantum coherence effect of the
generated chiral cat states, we examine the Wigner function of the generate states. To evaluate the validity of the approximation, additionally, we derive the exact states governed by the full Hamiltonian~(\ref{eq2}) and evaluate the fidelities between the approximate states and the exact states.

\subsection{Analytical results of the chiral cat states and their Wigner functions governed by the approximate Hamiltonian $H_{\mathrm{app}}$}
The unitary evolution operator associated with the approximate Hamiltonian $%
H_{\mathrm{app}}$ is given by $U_{\mathrm{app}}(t)=\mathrm{exp}(-iH_{\mathrm{app}}t)$. To generate the chiral cat states, we assume that the system is initially prepared in the state
\begin{equation}\label{eq8}
\left\vert \psi _{\mathrm{app}}\left( 0\right) \right\rangle =\frac{1}{\sqrt{%
2}}\left( \left\vert g\right\rangle +\left\vert e\right\rangle \right)
\left\vert \alpha \right\rangle _{\mathrm{CW}}\left\vert \alpha
\right\rangle _{\mathrm{CCW}},
\end{equation}
where $\vert \alpha \rangle _{\mathrm{CW}}$ and $\vert \alpha \rangle _{\mathrm{CCW}}$ denote the coherent states of the CW and CCW modes, respectively. The state of the system at time $t$ can be obtained as
\begin{eqnarray}\label{eq9}
\vert \psi _{\mathrm{app}}( t) \rangle  &\!=\!&\frac{1}{%
\sqrt{2}}[e^{-i\vartheta (t)}|g\rangle \vert \alpha
_{\mathrm{CW}}^{(+)}(t)\rangle _{\mathrm{CW}}\vert \alpha _{\mathrm{CCW}}^{(+)}(t)\rangle _{%
\mathrm{CCW}}  \nonumber   \\
&&+e^{i\vartheta (t)}|e\rangle \vert \alpha_{\mathrm{CW}}^{(-)}(t)\rangle _{\mathrm{CW}}\vert \alpha_{\mathrm{CCW}}^{(-)}(t)\rangle _{%
\mathrm{CCW}}],
\end{eqnarray}
where we introduce the phase factor $\vartheta (t)=(\zeta_{\mathrm{CW}}+%
\zeta_{\mathrm{CCW}})t/2$ and the displacement amplitudes $\alpha_{\epsilon=\mathrm{CW,CCW}}^{(\pm)}(t) = \alpha e^{-i(\Delta_{\mathrm{\epsilon}}\pm \zeta_{\mathrm{\epsilon}})t} $
for the CW and CCW modes. The values of these coherent state amplitudes $\alpha_{\mathrm{CW}}^{(\pm)}(t)$ and $\alpha_{\mathrm{CCW}}^{(\pm)}(t)$ are crucial for the generation of the chiral cat states. Specifically, for the same mode (denoted by the same subscripts), the difference between $\alpha^{(+)}(t)$ and $\alpha^{(-)}(t)$ promotes the generation of the cat states. Conversely, when the superscripts are identical, the variation in amplitudes between the CW and CCW modes induces the evolution of distinct quantum states, thereby driving the dynamics of the chiral cat states.

To generate quantum superposed states of the traveling modes, we measure the state of the atom using the basis states $|\pm \rangle =(|e\rangle \pm |g\rangle)/\sqrt{2}$. By expressing the atomic state with the bases $|\pm \rangle$, the generated state $\left\vert \psi _{\mathrm{app}}\left( t\right) \right\rangle$ can be reexpressed as
\begin{equation}\label{eq12}
\left\vert \psi _{\mathrm{app}}\left( t\right) \right\rangle =\frac{1}{2%
\mathcal{M}_{+}\left( t\right) }|+\rangle |\psi _{+}(t)\rangle +\frac{1}{2%
\mathcal{M}_{-}\left( t\right) }|-\rangle |\psi _{-}(t)\rangle,
\end{equation}
where we introduce
\begin{eqnarray}\label{eq13}
|\psi _{\pm }( t) \rangle  &=&\mathcal{M}_{\pm }( t)
[e^{i\vartheta (t)}\vert \alpha _{\mathrm{CW}%
}^{(-)}(t)\rangle _{\mathrm{CW}}\vert \alpha _{\mathrm{CCW}%
}^{(-)}(t)\rangle _{\mathrm{CCW}}  \nonumber \\
&&\pm e^{-i\vartheta (t)}\vert \alpha _{\mathrm{CW}%
}^{(+)}(t)\rangle _{\mathrm{CW}}\vert \alpha _{\mathrm{CCW}%
}^{(+)}(t)\rangle _{\mathrm{CCW}}],
\end{eqnarray}
with the normalization constants
\begin{eqnarray}\label{eq14}
\mathcal{M}_{\pm }( t)  &=&\Biggl[2\pm \sum_{l =1,-1}\exp \Biggl[2l
i\vartheta (t)+\vert \alpha \vert ^{2}  \nonumber \\
&&\times \Biggl(\sum_{\epsilon =\mathrm{cw},\mathrm{ccw%
}}\mathrm{exp}(2li\zeta_{\epsilon }t)-2\Biggl)\Biggl]\Biggl]^{-\frac{1}{2}}.
\end{eqnarray}
The measurement probabilities corresponding to the detected atomic states $|\pm \rangle$ are
\begin{eqnarray}\label{eq15}
\mathcal{P}_{\pm }( t)&=&\frac{1}{4}\Biggl[2\pm \sum_{l =1,-1}\exp \Biggl[2l
i\vartheta (t)+\vert \alpha \vert ^{2}  \nonumber \\
&&\times \Biggl(\sum_{\epsilon =\mathrm{cw},\mathrm{ccw%
}}\mathrm{exp}(2li\zeta_{\epsilon }t)-2\Biggl)\Biggl]\Biggl].
\end{eqnarray}
We will compare these probabilities given in Eq.~(\ref{eq15}) with those results corresponding to the exact state case in the next subsection.

Typically, the states $|\psi_{\pm}(t)\rangle$ of the CW and CCW modes are entangled states. However, when the two coherent state amplitudes for one of the two modes are overlapped, i.e., $\alpha^{(+)}_{\mathrm{CW}/\mathrm{CCW}}=\alpha^{(-)}_{\mathrm{CW}/\mathrm{CCW}}$, then the two modes will be disentangled. For example, we set the effective detunings $\Delta_{\mathrm{CW}}=2\Delta_{\mathrm{CCW}}$; then the CW mode is in a maximally distinguishable coherent state (cat state) while the CCW mode can remain in a coherent state at time $t_{s}=\pi/(2\zeta_{\mathrm{CW}})$. We denote this phenomenon as the generation of chiral cat states, i.e., creating distinct cat states in the two different rotating modes of the cavity. Note that here the coherent state can be understood as a special case of the cat state, with a zero distance between the two superposition components in phase space, which can mostly exhibit the contrast with the cat state. The two modes of the resonator can also be prepared in two distinct cat states by choosing other measurement times for the atom. When $\Delta_{\mathrm{CW}}\neq2\Delta_{\mathrm{CCW}}$,  it becomes impossible to simultaneously achieve a coherent state and a maximally distinguishable cat state. In this case, one mode is in a coherent state, while the other mode is in a less distinguishable cat state. Nonetheless, a chiral cat state can still be generated under these conditions.

\begin{figure}[tbp]
\centering \includegraphics[width=0.48\textwidth]{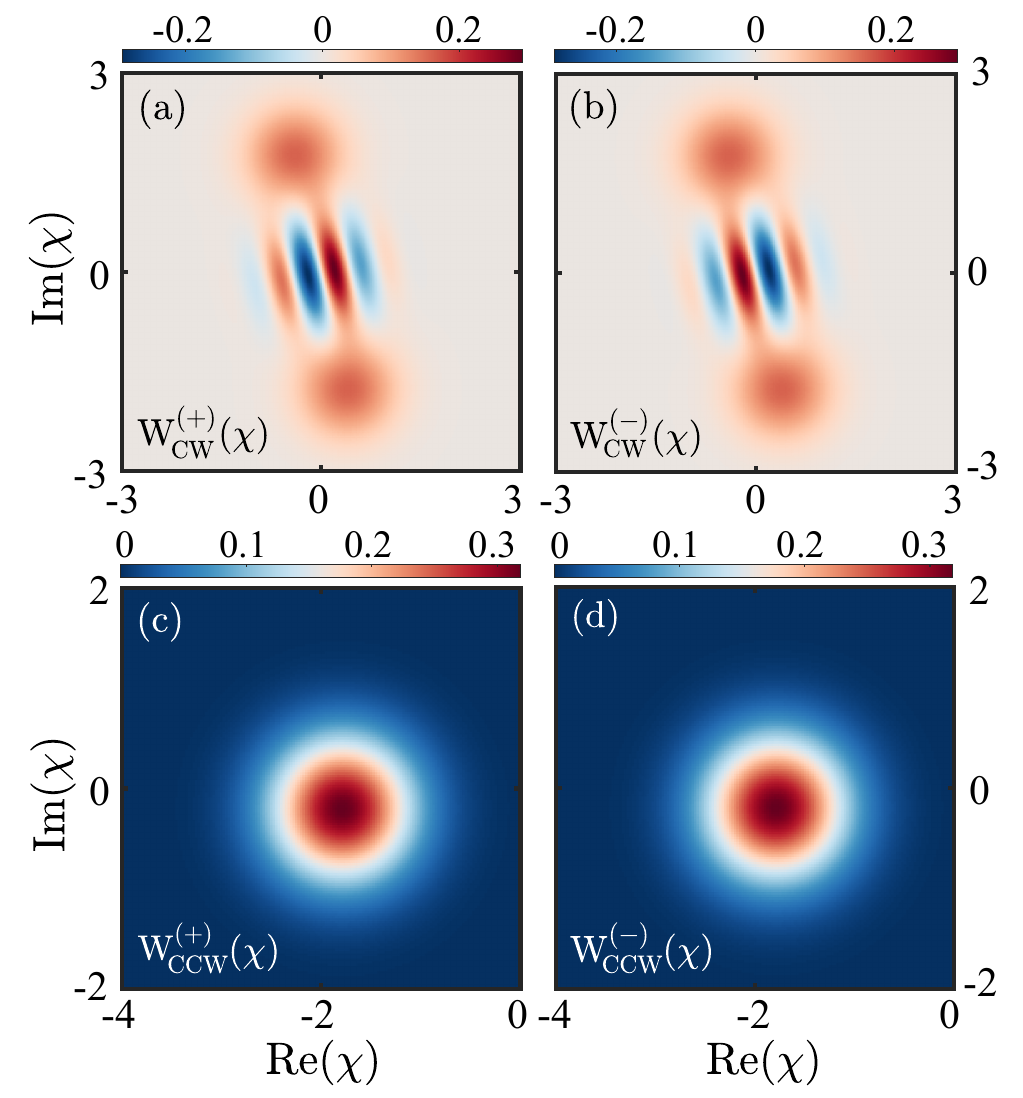}
\caption{Wigner functions $\mathrm{W}_{\mathrm{CW,CCW}}^{(\pm )}(\chi)$ of the generated states $\rho^{(\pm)}_{\mathrm{CW,CCW}}$ at time $t_{s}=\pi/(2\zeta_{\mathrm{CW}})$. Panels (a) and (b) [(c) and (d)] show the Wigner functions $\mathrm{W}_{\mathrm{CW}}^{(+)}(\chi)$ and $\mathrm{W}_{\mathrm{CW}}^{(-)}(\chi)$ [$\mathrm{W}_{\mathrm{CCW}}^{(+)}(\chi)$ and $\mathrm{W}_{\mathrm{CCW}}^{(-)}(\chi)$] for the CW (CCW) mode, respectively. The coherent amplitude of the initial coherent state $|\alpha\rangle$ is $\alpha=1.8$. Other parameters used are $\Delta/J=33$ and $\Delta_{\mathrm{sag}}/J=11$.}
\label{Fig2}
\end{figure}
To demonstrate the quantum coherence effect in the generated chiral cat states, we calculate the Wigner function~\cite{Barnett1997} for both the CW and CCW modes. When the CW and CCW modes are in the states described by the density matrices $\rho_{\mathrm{CW}}$ and $\rho_{\mathrm{CCW}}$, respectively, the Wigner functions are defined by
\begin{equation}\label{Winger}
\mathrm{W}_{\epsilon=\mathrm{CW,CCW}}(\chi)=\frac{2}{\pi}\mathrm{Tr}[D_{\epsilon}^{\dagger}(\chi)\rho_{\epsilon}D_{\epsilon}(\chi)(-1)^{a_{\epsilon}^{\dagger}a_{\epsilon}}],
\end{equation}
where $D_{\epsilon}(\chi)=\mathrm{exp}(\chi a_{\epsilon}^{\dagger}-\chi^{\ast} a_{\epsilon})$ is the displacement operator. For the states given in Eq.~(\ref{eq13}), we take the outer product $\rho^{(\pm)}=|\psi _{\pm }( t)\rangle\langle\psi _{\pm }( t) |$ and trace out one of the modes to obtain $\rho^{(\pm)}_{\epsilon}=\mathrm{Tr}_{\bar{\epsilon}}(\rho^{(\pm)})$, where $\bar{\epsilon}$ denotes the complement of $\epsilon$, i.e., $\overline{\mathrm{CCW}}=\mathrm{CW}$ and $\overline{\mathrm{CW}}=\mathrm{CCW}$. In terms of these reduced density matrices, the Wigner
functions $\mathrm{W}^{(\pm)}_{\mathrm{CW,CCW}}(\chi)$ can be obtained as
\begin{eqnarray}
\mathrm{W}_{\epsilon =\mathrm{CW,CCW}}^{(\pm )}( \chi ) &=&\frac{2\vert
\mathcal{M}_{\pm }\vert ^{2}}{\pi }[\Lambda_{\epsilon}^{(1)}(\chi)\pm\Lambda_{\epsilon}^{(2)}(\chi)\notag \\ &&\pm\Lambda_{\epsilon}^{(3)}(\chi)+\Lambda_{\epsilon}^{(4)}(\chi)],
\end{eqnarray}
where we introduce
\begin{subequations}
\begin{eqnarray}
\Lambda_{\epsilon}^{(1)}(\chi) \!\!&=\!\!& \exp [ 2[ \alpha _{\epsilon }^{(+)}(t)] ^{\ast }\chi +2\chi
^{\ast }\alpha _{\epsilon }^{(+)}(t)-2\vert \alpha \vert
^{2}-2\vert \chi \vert ^{2}],\notag \\ \\
\Lambda_{\epsilon}^{(2)}(\chi)\!\! &=\!\!& \exp [ -2i\vartheta ( t) +\vert \alpha \vert
^{2}( e^{-2i\zeta _{\bar{\epsilon}}t}-e^{-2i\zeta _{\epsilon
}t}-2) \notag \\ &&+2\chi [ \alpha _{\epsilon }^{(-)}(t)] ^{\ast } +2\chi
^{\ast }\alpha _{\epsilon }^{(+)}(t)-2\vert \chi \vert ^{2}], \\
\Lambda_{\epsilon}^{(3)}(\chi)\! \!&=\!\!& \exp [ 2i\vartheta ( t) +\vert \alpha \vert
^{2}( e^{2i\zeta _{\bar{\epsilon}}t}-e^{2i\zeta _{\epsilon
}t}-2) \notag \\ && +2\chi [ \alpha _{\epsilon }^{(+)}(t)] ^{\ast }+2\chi
^{\ast }\alpha _{\epsilon }^{(-)}(t)-2\vert \chi \vert ^{2}], \\
\Lambda_{\epsilon}^{(4)}(\chi) \!\!&=\!\!& \exp [ 2[ \alpha _{\epsilon }^{(-)}(t)] ^{\ast }\chi +2\chi
^{\ast }\alpha _{\epsilon }^{(-)}(t)-2\vert \alpha \vert
^{2}-2\vert \chi \vert ^{2}]\notag. \\
\end{eqnarray}
\end{subequations}

In Figs.~\ref{Fig2}(a)-\ref{Fig2}(d), we show the Wigner functions $\mathrm{W}_{\mathrm{CW,CCW}}^{(\pm )}(\chi)$ for both the CW and CCW modes at time $t_{s}=\pi/(2\zeta_{\mathrm{CW}})=69.11J^{-1}$.~Concretely, Figs.~\ref{Fig2}(a) and~\ref{Fig2}(b) display the Wigner functions $\mathrm{W}_{\mathrm{CW}}^{(+)}(\chi)$ and $\mathrm{W}_{\mathrm{CW}}^{(-)}(\chi)$ for the CW mode, respectively. Here we can see distinct interference fringes caused by quantum superposition of two coherent states. The interference fringes indicate the presence of coherent superposition states, commonly referred to as cat states. In contrast, the Wigner functions for the CCW mode exhibit Gaussian peaks in the phase space, as shown in Figs.~\ref{Fig2}(c) and~\ref{Fig2}(d), which are signature for coherent states in the CCW mode. We note that the appearance of the coherent state for the CCW mode is caused by the overlap between the two coherent amplitudes $\alpha _{\mathrm{CCW}}^{(-)}(t_{s})$ and $\alpha _{\mathrm{CCW}}^{(+)}(t_{s})$. Comparing Figs.~\ref{Fig2}(a) [\ref{Fig2}(b)] with \ref{Fig2}(c) [\ref{Fig2}(d)], we observe that the CW mode is in a cat state, while the CCW mode is in a coherent state, demonstrating the generation of chiral cat states.

\subsection{Exact states determined by the full Hamiltonian $H$ and the fidelities between the exact and approximate states}
\begin{figure}[tbp]
\centering \includegraphics[width=0.48\textwidth]{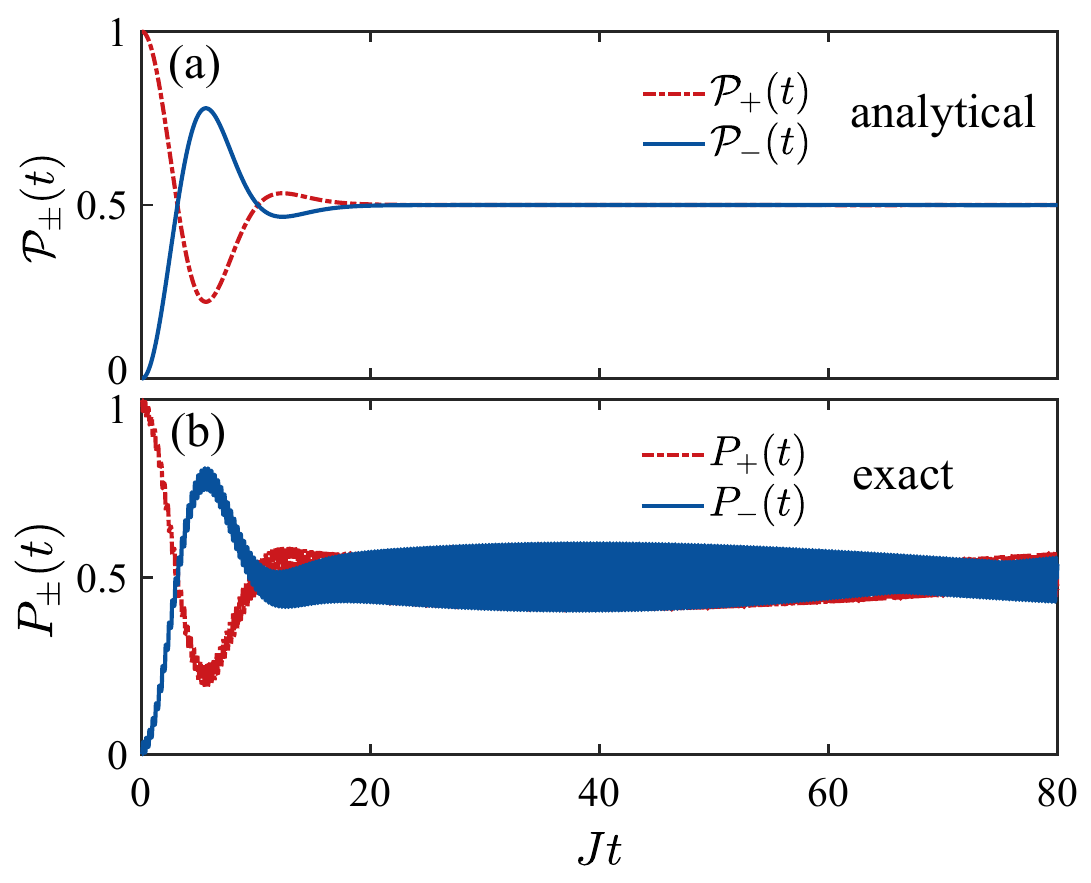}
\caption{Analytical measurement probabilities $\mathcal{P}_{\pm}(t)$ (a) from Eq.~(\ref{eq15}) and the exact measurement probabilities $P_{\pm}(t)$ (b) from Eq.~(\ref{eq26}) as functions of the scaled evolution time $Jt$. For the exact solution in panel (b), both the CW and CCW modes have a truncation dimension of 13. Other parameters used in panels (a) and (b) are the same as those in Fig.~\ref{Fig2}.}
\label{Fig3}
\end{figure}
In the above section, we have calculated the generated states governed by the approximate Hamiltonian. To evaluate the state generation in a realistic case, we need to calculate the state generation with the full Hamiltonian, and evaluate the fidelity between the approximate and exact states. The exact state of this system can be determined by numerically solving the Schr\"{o}dinger equation with the full Hamiltonian~(\ref{eq2}). For the present system, a general pure state can be expressed as
\begin{eqnarray}\label{eq22}
\vert \Psi ( t) \rangle &=&\sum_{m,k=0}^{\infty }[
A_{m,k}( t)\vert e\rangle \vert m\rangle
_{\mathrm{CW}}\vert k\rangle _{\mathrm{CCW}}\notag \\ &&+B_{m,k}( t) \vert
g\rangle\vert m\rangle _{\mathrm{CW}}\vert k\rangle
_{\mathrm{CCW}}],
\end{eqnarray}%
where $\vert m\rangle
_{\mathrm{CW}}$ and $\vert k\rangle _{\mathrm{CCW}}$ represent the Fock states of the CW and CCW modes, respectively. $A_{m,k}( t)$ and $B_{m,k}( t)$ are the probability amplitudes. By substituting Eq.~(\ref{eq2}) and Eq.~(\ref{eq22}) into the Schr\"{o}dinger equation, we derive the equations of motion for the probability amplitudes $A_{m,k}( t)$ and $B_{m,k}( t)$ as
\begin{subequations}
\begin{eqnarray}\label{eq23a}
\dot{A}_{m,k}\left( t\right) &=& -i\left( m\Delta _{\mathrm{CW}}+k\Delta _{\mathrm{CCW}}\right)
A_{m,k}\left( t\right)-iJ\sqrt{m+1} \notag\\&&\times B_{m+1,k}\left( t\right) -iJ\sqrt{k+1}B_{m,k+1}\left( t\right) ,\\ \label{eq23b}
\dot{B}_{m,k}\left( t\right) &=&-i\left( m\Delta _{\mathrm{CW}}+k\Delta _{\mathrm{CCW}}\right)
B_{m,k}\left( t\right) -iJ\sqrt{m}
\notag\\&&\times A_{m-1,k}\left( t\right) -iJ\sqrt{k%
}A_{m,k-1}\left( t\right).
\end{eqnarray}%
\end{subequations}
For the initial state $|+\rangle\left\vert \alpha \right\rangle _{\mathrm{CW}}\left\vert \alpha
\right\rangle _{\mathrm{CCW}}$, the initial conditions for the probability amplitudes are $A_{m,k}( 0) =B_{m,k}( 0) =e^{-
\left\vert \alpha\right\vert ^{2} }\alpha^{m+k}/\sqrt{2m!k!}$. Using these initial conditions, the evolution of the probability amplitudes $A_{m,k}( t)$ and $B_{m,k}( t)$ can be obtained by numerically solving Eqs.~(\ref{eq23a}) and~(\ref{eq23b}), then the exact state of the system is obtained.

When the atom is detected in the states $|\pm\rangle$, the CW and CCW modes collapse into the corresponding states
\begin{equation}\label{eq25}
|\Psi_{\pm}(t)\rangle\!\!=\!\!\frac{1}{\sqrt{2P_{\pm}(t)}}\!\!\sum_{m,k=0}^{\infty}[A_{m,k}( t)\pm B_{m,k}( t)] \vert m\rangle
_{\mathrm{CW}}\vert k\rangle _{\mathrm{CCW}},
\end{equation}
with the corresponding detection probabilities
\begin{equation}\label{eq26}
P_{\pm}(t)=\frac{1}{2}\sum_{m,k=0}^{\infty}|[A_{m,k}( t)\pm B_{m,k}( t)] |^{2}.
\end{equation}

In Fig.~\ref{Fig3}(a), we display the analytical  measurement probabilities $\mathcal{P}_{\pm}(t)$, given by Eq.~(\ref{eq15}), vs the scaled evolution time $Jt$. We observe that the sum of $\mathcal{P}_{+}(t)$ and $\mathcal{P}_{-}(t)$ is normalized. Furthermore, for longer subsequent measurement time $t_{m}$, the measurement  probabilities $\mathcal{P}_{\pm}(t)\approx1/2$, corresponding to $\sum_{l =1,-1}\exp [2l
i\vartheta (t_{m})+\left\vert \alpha \right\vert ^{2} [\sum_{\epsilon =\mathrm{CW},\mathrm{CCW%
}}\mathrm{exp}(2li\zeta_{\epsilon }t_{m})-2]]\rightarrow0$ in Eq.~(\ref{eq15}). The exact measurement  probabilities $P_{\pm}(t)$ given by Eq.~(\ref{eq26}) are shown in Fig.~\ref{Fig3}(b). The exact  measurement  probabilities $P_{\pm}(t)$ exhibit oscillating behavior around 1/2, which is caused by the last term in Eq.~(\ref{eq4}). By comparing the analytical results with the numerical results, we find that the analytical probability evolution is smooth, while the numerical results exhibit oscillation. Nevertheless, the patterns of the approximate and exact results agree well, which  confirms the validity of the approximate made during the derivation of the approximate Hamiltonian.

To quantitatively verify the validity of the approximate Hamiltonian $H_{\mathrm{app}}$, we evaluate the fidelity $F(t)=|\langle\Psi ( t)\vert \psi _{\mathrm{app}}( t) \rangle|^{2}$ between the exact and approximate states. Here, $\vert \psi _{\mathrm{app}}( t) \rangle$ represents the approximate state governed by the approximate Hamiltonian $H_{\mathrm{app}}$ in Eq.~(\ref{eq6}) and $\vert \Psi ( t) \rangle$ represents the exact state governed by the full Hamiltonian $H$ in Eq.~(\ref{eq2}). Based on Eqs.~(\ref{eq9}) and~(\ref{eq22}), the fidelity can be calculated as
\begin{eqnarray}\label{Fieq1}
F(t)&\!\!=\!\!&\Biggl\vert e^{-\left\vert \alpha \right\vert
^{2}}\sum_{m,k=0}^{\infty }\Biggl[
e^{i\vartheta (t) }A_{m,k}^{\ast }( t) \frac{[
\alpha _{\mathrm{CW}}^{(-)}( t) ] ^{m}[ \alpha _{\mathrm{%
CCW}}^{(-)}(t)] ^{k}}{\sqrt{2m!k!}} \notag \\ && +
e^{-i\vartheta ( t) }B_{m,k}^{\ast }( t) \frac{\left[
\alpha _{\mathrm{CW}}^{(+)}( t) \right] ^{m}[ \alpha _{\mathrm{%
CCW}}^{(+)}(t)] ^{k}}{\sqrt{2m!k!}}%
\Biggl] \Biggl\vert ^{2}.
\end{eqnarray}
Similarly, we can also evaluate the performance of the chiral-cat-states generation by calculating the fidelities between the generate states $|\Psi_{\pm}(t)\rangle$ in Eq.~(\ref{eq25}) and the target states $|\psi_{\pm}(t)\rangle$ in Eq.~(\ref{eq13}). Using Eqs.~(\ref{eq25}) and~(\ref{eq13}), the fidelities $F_{\pm}(t)=|\langle\Psi _{\pm}( t)\vert \psi _{\pm}( t) \rangle|^{2}$ can be obtained as
\begin{eqnarray}\label{Fieq2}
F_{\pm}(t)&=&\frac{|\mathcal{M}_{\pm}(t)|^{2}e^{-2\vert \alpha \vert^{2}}}{P_{\pm}(t)}\Biggl\vert \sum_{m,k=0}^{\infty }\Biggl[(A_{m,k}^{\ast }( t)\pm B_{m,k}^{\ast }( t) ) \notag \\ &&\times\Biggl(
e^{i\vartheta( t) } \frac{\Bigl[
\alpha _{\mathrm{CW}}^{(-)}( t) \Bigl] ^{m}\Bigl[ \alpha _{\mathrm{%
CCW}}^{(-)}(t)\Bigl] ^{k}}{\sqrt{2m!k!}}  \notag \\ && \pm
e^{-i\vartheta ( t) }\frac{\Bigl[
\alpha _{\mathrm{CW}}^{(+)}( t) \Bigl] ^{m}\Bigl[ \alpha _{\mathrm{%
CCW}}^{(+)}(t)\Bigl] ^{k}}{\sqrt{2m!k!}}%
\Biggl) \Biggl]\Biggl\vert ^{2}.
\end{eqnarray}

\begin{figure}[tbp]
\centering \includegraphics[width=0.48\textwidth]{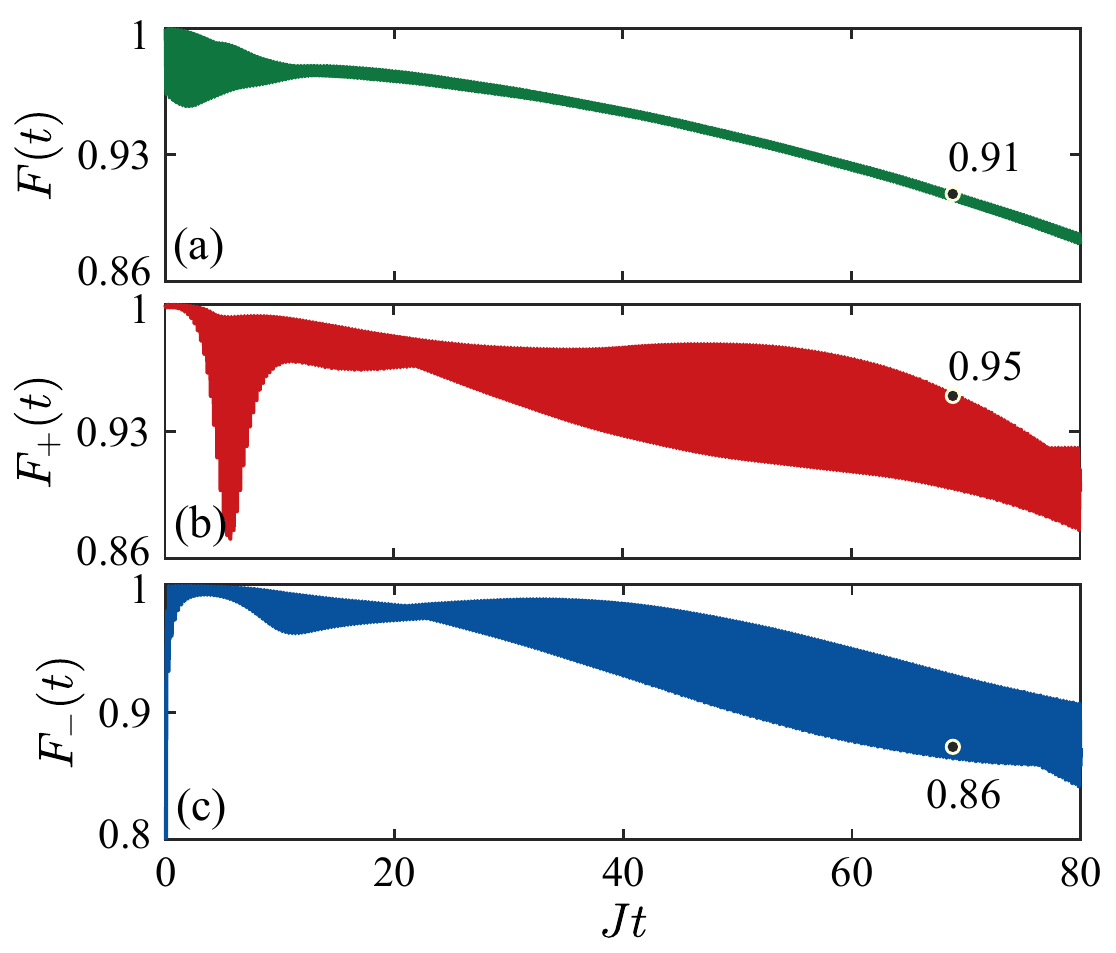}
\caption{Fidelities (a) $F(t)$, (b) $F_{+}(t)$, and (c) $F_{-}(t)$ as functions of the scaled evolution time $Jt$. The parameters used are the same as those in Fig.~\ref{Fig2}.}
\label{Fig4}
\end{figure}
Figure~\ref{Fig4} illustrates the fidelities $F(t)$ and $F_{\pm}(t)$ as functions of the scaled evolution time $Jt$. We observe that the fidelities $F(t)$ and $F_{\pm}(t)$ decrease gradually and exhibit oscillations over time. The decreasing fidelity over time indicates a gradual loss of correspondence between the approximate and exact states. From Figs.~\ref{Fig4}(a) to~\ref{Fig4}(c), we see that at the target moment $t_{s}=69.11J^{-1}$, the fidelities $F(t_{s})$, $F_{+}(t_{s})$, and $F_{-}(t_{s})$ take the values $0.91$, $0.95$, and $0.86$, respectively. These relatively high fidelity values indicate that the approximate states and the exact states match well at the measurement time, demonstrating the validity of the approximation. It is worth noting that higher fidelity values can be achieved by adjusting the system parameters. However, due to the relevance between these parameters, achieving a higher fidelity usually requires a longer target time. In the open-system case, a longer evolution time will lead to a stronger cavity-photon dissipation, thereby reducing the fidelity. In realistic cases, we need to choose proper measurement time to reach a proper balance between the fidelity and the dissipation.

\subsection{Wigner functions of the exact states}
\begin{figure}[tbp]
\centering \includegraphics[width=0.48\textwidth]{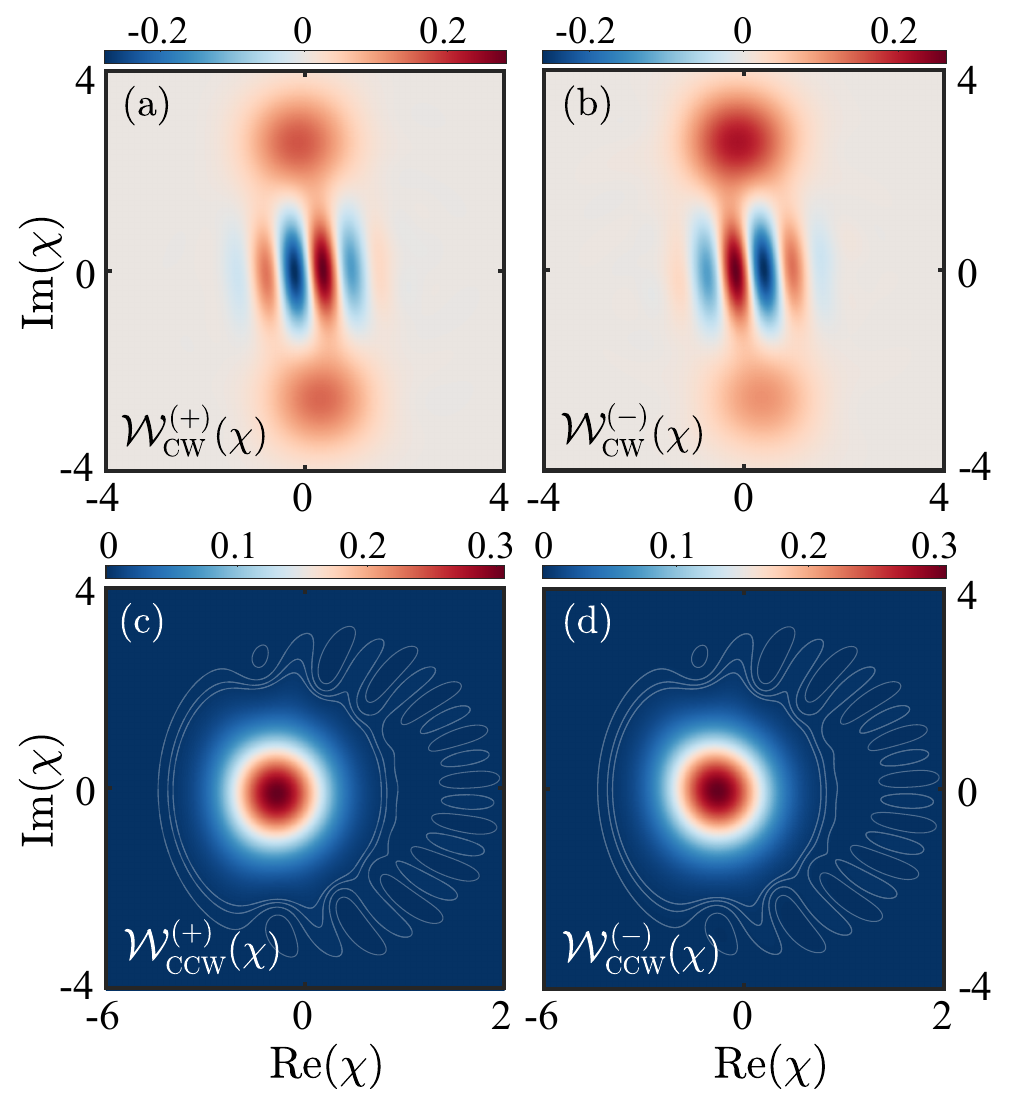}
\caption{(a)$-$(d) Wigner functions $\mathcal{W}^{(\pm)}_{\mathrm{CW,CCW}}(\chi)$ for the exact states $|\Psi_{\pm}(t_{s})\rangle$  in Eq.~(\ref{eq25}) at time $t_{s}=\pi/(2\zeta_{\mathrm{CW}})$. Other parameters used are the same as those in Fig.~\ref{Fig2}.}
\label{Fig5}
\end{figure}
To confirm the chiral state generation in the phase space, we calculate the Wigner functions of the exact states $|\Psi_{\pm}(t)\rangle$. Following the preceding discussions and using Eqs.~(\ref{Winger}) and~(\ref{eq25}), we derive the Wigner functions $\mathcal{W}^{(\pm)}_{\mathrm{CW,CCW}}(\chi)$ for the exact states $|\Psi_{\pm}(t)\rangle$ as
\begin{subequations}
\begin{eqnarray}
\mathcal{W}_{\mathrm{CW}}^{(\pm) }( \chi ) &\!\!=\!\!&  \frac{1}{\pi P_{\pm
}( t) }\sum\limits_{l,m,n,k=0}^{\infty }( -1) ^{l}%
[ A_{m,k}( t) +B_{m,k}( t) ]   \notag \\
&&\times [ A_{n,k}^{\ast }( t) +B_{n,k}^{\ast }(
t) ]\mathcal{L}_{m,l}^{\mathrm{CW}\dagger}(\chi)\mathcal{L}_{n,l}^{\mathrm{CW}}(\chi) ,  \notag \\  \\
\mathcal{W}_{\mathrm{CCW}}^{(\pm) }( \chi ) \! &=& \! \frac{1}{\pi P_{\pm
}( t) }\sum\limits_{l,m,k,j=0}^{\infty }( -1) ^{l}%
[ A_{m,k}( t) +B_{m,k}( t) ]   \nonumber \\
&&\times [ A_{m,j}^{\ast }( t) +B_{m,j}^{\ast }(
t) ] \mathcal{L}_{k,l}^{\mathrm{CCW}\dagger}(\chi)\mathcal{L}^{\mathrm{CCW}}_{j,l}(\chi).\nonumber \\
\end{eqnarray}
\end{subequations}
Here the measurement probabilities $P_{\pm}( t)$ are given by Eq.~(\ref{eq26}), and $\mathcal{L}^{\epsilon=\mathrm{CW},\mathrm{CCW}}_{n,l}(\chi)$ is defined as $\mathcal{L}_{n,l}^{\epsilon}(\chi)={}_{\epsilon}\langle n \vert D( \chi ) \vert l \rangle_{\epsilon}$, which is given by~\cite{Oliveira1990}
\begin{equation}
\begin{aligned}\label{663}
\mathcal{L}_{n,l}^{\epsilon}(\chi) =
\begin{cases}
\sqrt{\frac{n!}{l!}} e^{-|\chi|^2/2} (-\chi^*)^{l-n} L_n^{l-n} (|\chi|^2), & l > n, \\
\sqrt{\frac{l!}{n!}} e^{-|\chi|^2/2} (\chi)^{n-l} L_l^{n-l} (|\chi|^2), & n > l,
\end{cases}
\end{aligned}
\end{equation}
where $L_n^{l-n} (x)$ are the associated Laguerre polynomials.

In Fig.~\ref{Fig5}, we show the exact Wigner functions $\mathcal{W}^{(\pm)}_{\mathrm{CW,CCW}}(\chi)$ at time $t_{s}=69.11J^{-1}$. Figures~\ref{Fig5}(a) and~\ref{Fig5}(b) show clear interference fringes for the CW mode under the full Hamiltonian, which match well with the approximate solutions in Figs.~\ref{Fig2}(a) and~\ref{Fig2}(b). This indicates that the approximation is valid in describing the nonclassical characteristics of the CW mode. For the CCW mode, however, the exact Wigner functions in Figs.~\ref{Fig5}(c) and~\ref{Fig5}(d) show small interference fringes that are not presented in the approximate solutions. These small fringes appear because the components $\alpha _{\mathrm{CCW}}^{(-)}(t)$ and $\alpha _{\mathrm{CCW}}^{(+)}(t)$ in phase space are not completely overlapped (the Wigner function does not form a perfectly circular Gaussian peak), resulting in some interference effects. Despite this, we can still clearly distinguish the CW and CCW modes under the full Hamiltonian, demonstrating the feasibility for generating chiral cat states with our scheme.

\section{Generation of chiral cat states in the Open-system case\label{sec4}}
In this section, we investigate the generation of chiral cat states in the open-system case. Specifically, we analyze in detail how the system dissipations affect the state-generation performance, namely, the fidelity and the state-generation probability. We also explore the influence of the dissipations on quantum coherence by calculating the Wigner functions of the generated cat states in the open-system case.

\subsection{Quantum master equation and its exact solution}
In real physical systems, the dissipations inevitably affect the dynamics of the system. For the present cavity-QED system, the average thermal excitation numbers for the cavity field and the atom are negligibly small; then we can assume that both the cavity fields and the atom are connected to individual vacuum baths. By introducing the dissipation terms for both the cavity modes and the atom, the dynamics of the system is described by the quantum master equation
\begin{equation}\label{eq29}
\dot{\rho}=-i[H,\rho]+\kappa L[a_{\mathrm{CW}}]\rho+\kappa L[a_{\mathrm{CCW}}]\rho+\gamma L[\sigma_{-}]\rho,
\end{equation}
where the Hamiltonian $H$ is given by Eq.~(\ref{eq2}). The parameters $\kappa$ and $\gamma$ are the decay rates of the cavity field and the two-level atom, resepectively. The superoperator $L[o]\rho=o\rho o^{\dagger}-(o^{\dagger}o\rho+\rho o^{\dagger}o)/2$ denotes the Lindbland term for the operators $o=a_{\mathrm{CW}}, a_{\mathrm{CCW}}$, and $\sigma_{-}$.

To solve the quantum master equation, we expand the density matrix of the system using the Fock-state bases as follows:
\begin{eqnarray}\label{eq30}
\rho(t)&=&\sum_{w,v=e,g}\Biggl[\sum_{m,k,n,j=0}^{\infty}\rho_{w,m,k,v,n,j}(t)\notag\\&&\times|w\rangle|m\rangle_{\mathrm{CW}}|k\rangle_{\mathrm{CCW}}\langle v|_{\mathrm{CW}}\langle n|_{\mathrm{CCW}}\langle j|\Biggl],
\end{eqnarray}
with $\rho_{w,m,k,v,n,j}=\langle w|_{\mathrm{CW}}\langle m|_{\mathrm{CCW}}\langle k|\rho |v\rangle n\rangle_{\mathrm{CW}}|j\rangle_{\mathrm{CCW}}$. To numerically solve these equations of motion for the matrix density-matrix elements, we choose the initial state of the system as $|+\rangle |\alpha \rangle_{\mathrm{CW}}|\alpha\rangle_{\mathrm{CCW}}$; then the corresponding initial values of these density matrix elements are given by $\rho_{e,m,k,e,n,j}(0)=\rho_{e,m,k,g,n,j}(0)=\rho_{g,m,k,e,n,j}(0)=\rho_{g,m,k,g,n,j}(0)=e^{-2|\alpha|^{2}}\alpha^{m+k}(\alpha^{*})^{n+j}/2\sqrt{m!n!k!j!}$. With the numerical solutions, we can track the evolution of the density matrix and investigate the physical properties of the generated states.

\subsection{Fidelities between the approximate and exact states in the open-system case}
\begin{figure}[tbp]
	\centering \includegraphics[width=0.48\textwidth]{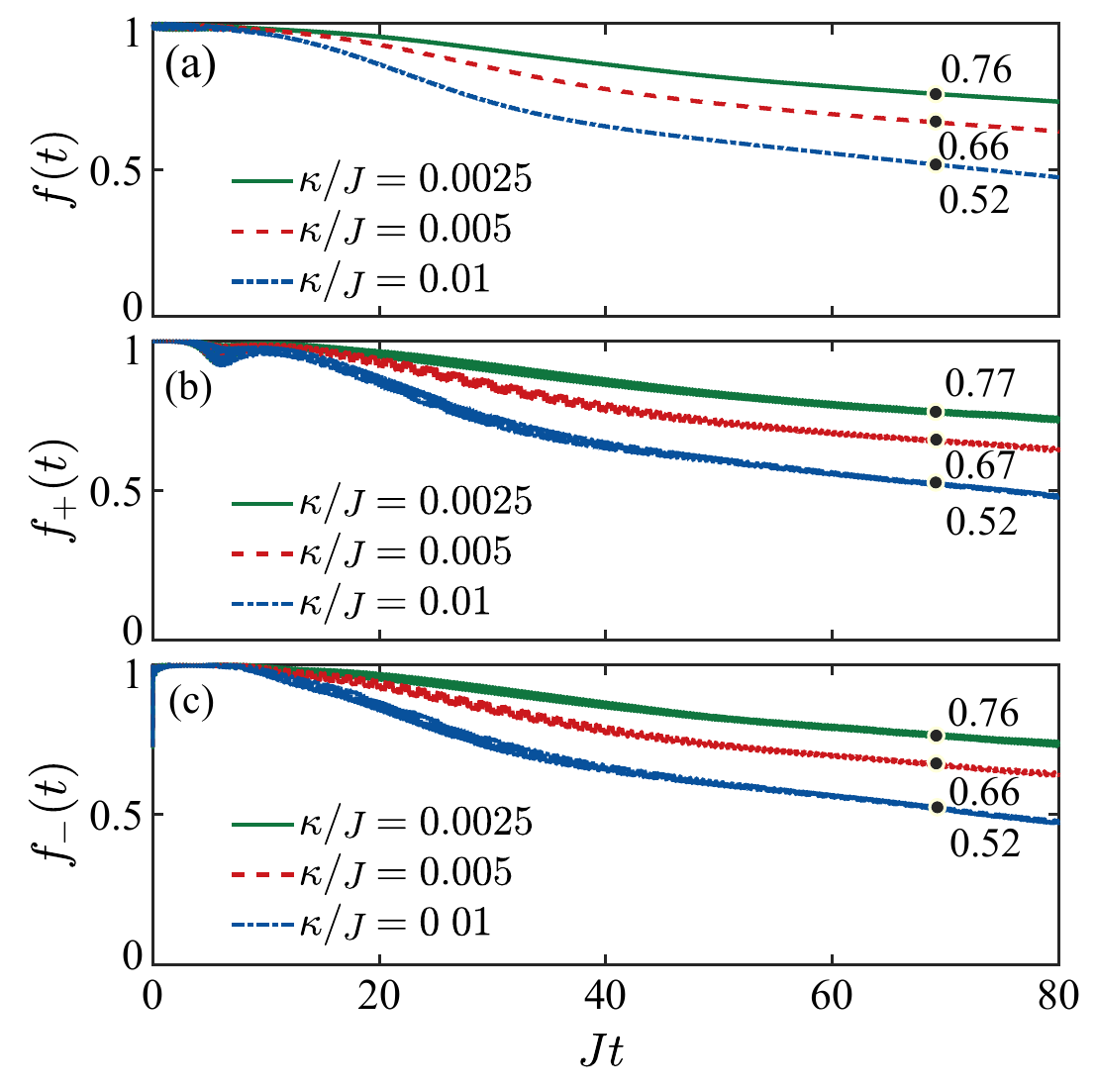}
	\caption{Fidelities (a) $f(t)$, (b) $f_{+}(t)$ , and (c) $f_{-}(t)$ as functions of the scaled evolution time $Jt$ for different cavity-field decay rates $\kappa/J=0.0025$, $0.005$, and $0.01$. The decay rate of the atom is $\gamma/J=0.005$. Other parameters used are the same as those in Fig.~\ref{Fig2}.}
	\label{Fig6}
\end{figure}
To investigate the effect of the dissipations on the state generation, we calculate the fidelity $f(t)=\langle\psi(t)|\rho(t)|\psi(t)\rangle$ between
the exact state $\rho(t)$ given in Eq.~(\ref{eq29}) and the analytical state $|\psi(t)\rangle$ given in Eq.~(\ref{eq9}). Particularly, for generating chiral cat states, a projective measurement upon the atomic states $|\pm\rangle$ is required. Then the corresponding density matrices for the CW and CCW modes become
\begin{eqnarray}\label{eq33}
\rho^{(\pm)}(t) &=& \frac{1}{2p_{\pm}(t)}\sum_{m,k,n,j}^{\infty} \Xi_{m,k,n,j}^{(\pm)}(t) \nonumber \\
&&\times |m\rangle_{\mathrm{CW}} |k\rangle_{\mathrm{CCW}} \,_{\mathrm{CW}}\langle n| \,_{\mathrm{CCW}}\langle j|,
\end{eqnarray}
where we introduce the variables
$\Xi_{m,k,n,j}^{(\pm)}(t)=\rho_{e,m,k,e,n,j}(t)+\rho_{g,m,k,g,n,j}(t)\pm\rho_{e,m,k,g,n,j}(t)\pm\rho_{g,m,k,e,n,j}(t)$
and the measurement probabilities
\begin{eqnarray}\label{eq34}
p_{\pm}(t)=\frac{1}{2}\sum_{m,k=0}^{\infty}\Xi_{m,k,m,k}^{(\pm)}(t)
\end{eqnarray}
corresponding to the atomic states $|\pm\rangle$. Subsequently, the fidelities $f_{\pm}(t)=\langle\psi_{\pm}(t)|\rho^{(\pm)}(t)|\psi_{\pm}(t)\rangle$ between the generated states $\rho^{(\pm)}(t)$ and the target states $|\psi_{\pm}(t)\rangle$ can be calculated accordingly.

\begin{figure}[tbp]
	\centering \includegraphics[width=0.48\textwidth]{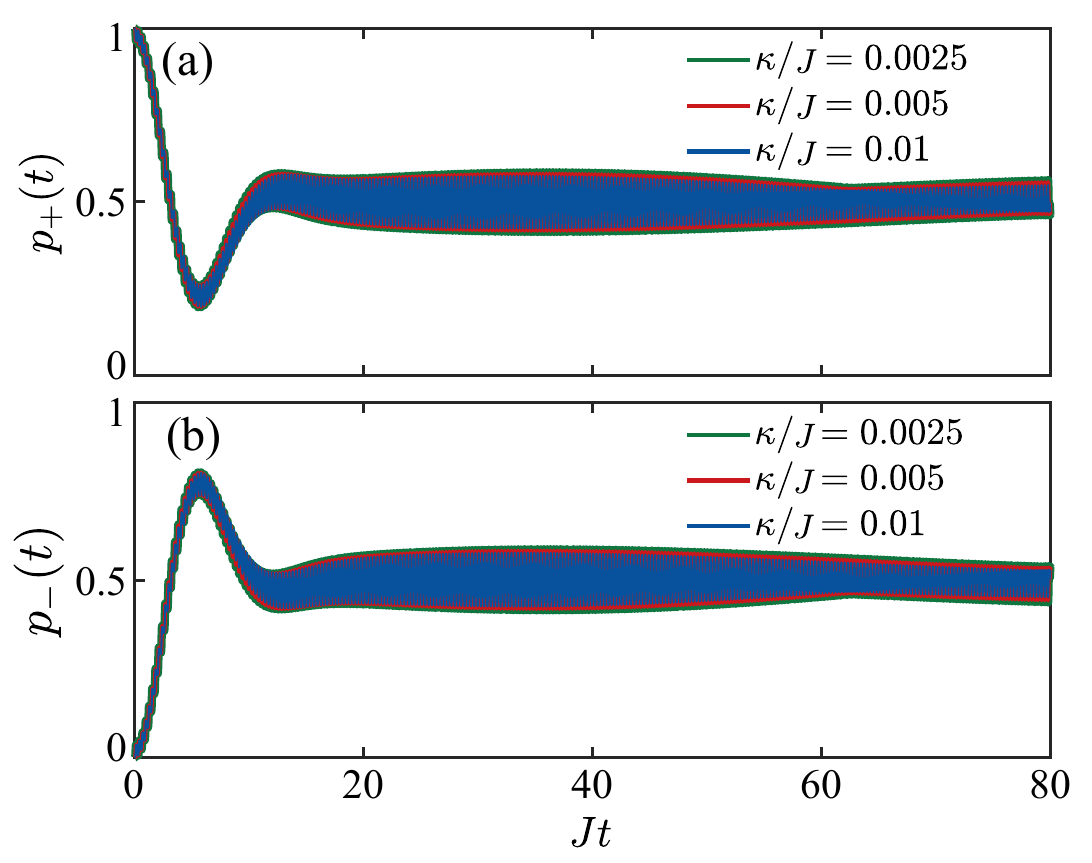}
	\caption{Measurement probabilities (a) $p_{+}(t)$ and (b) $p_{-}(t)$ given in Eq.~(\ref{eq34}) versus the scaled evolution time $Jt$ for different cavity-field decay rates $\kappa/J=0.0025$, $0.005$, and $0.01$ in the open-system case. The parameters used are the same as those in Fig.~\ref{Fig6}.}
	\label{Fig7}
\end{figure}
In Fig.~\ref{Fig6}, we show the fidelities $f(t)$ and $f_{\pm}(t)$ versus the scaled evolution time $Jt$ in the open-system case. Compared to the closed-system case in Fig.~\ref{Fig4}, the presence of system dissipation accelerates the decay of the fidelities and reduces oscillations. Specifically, increasing the cavity-field decay rate from $\kappa/J=0.0025$ to $0.005$ and $0.01$ leads to a more rapid fidelity reduction, as shown by the green solid, red dashed, and blue dotted curves, respectively. This trend is consistent for $f(t)$ and $f_{\pm}(t)$. Moreover, the results show that, at time $t_{s}=69.11J^{-1}$, the fidelities $f(t_{s})$, $f_{+}(t_{s})$, and $f_{-}(t_{s})$ take the values 0.76, 0.77, and 0.76 for $\kappa/J=0.0025$, 0.66, 0.67, and 0.66 for $\kappa/J=0.005$, and 0.52, 0.52, and 0.52 for $\kappa/J=0.01$, respectively. These results show that larger decay rates lead to reduced fidelity, indicating the degression of quantum state generation by the dissipations in the open-system case.

In Fig.~\ref{Fig7}, we show the measurement probabilities $p_{+}(t)$ and $p_{-}(t)$ versus the scaled evolution time $Jt$ corresponding to different cavity-field decay rates in the open-system case. Both the probabilities $p_{+}(t)$ [Fig.~\ref{Fig7}(a)] and $p_{-}(t)$ [Fig.~\ref{Fig7}(b)] oscillate around $0.5$. We observe that
the larger the dissipation rate is, the lower the amplitude of the oscillation envelope for the probabilities is (as depicted by the green, red, and blue curves for $\kappa/J=0.0025, 0.005$, and $0.01$, respectively). The reason for the reduction in the oscillation amplitude is that a stronger dissipation will cause more photons to decay, thereby reducing the impact of the last term in Eq.~(\ref{eq4}).

\subsection{The Wigner functions of the generated states in the open-system case}
In the open-system case, the dissipations of the system will reduce the quantum coherence in the generated chiral cat states. Using Eqs.~(\ref{Winger}) and~(\ref{eq33}), we can derive the Wigner functions $\tilde{\mathcal{W}}^{(\pm)}_{\mathrm{CW,CCW}}(\chi)$ in the open-system case as
\begin{subequations}
\begin{eqnarray}
\mathcal{\tilde{W}}_{\mathrm{CW}}^{( \pm ) }( \chi ) &=&%
\frac{1}{\pi p_{\pm }( t) }\sum\limits_{l,m,n,k=0}^{\infty }(-1)^{l}\Xi
_{m,k,n,k}^{( \pm ) }(t) \notag \\ && \times\mathcal{L}_{m,l}^{\mathrm{CW}\dagger}(\chi)\mathcal{L}_{n,l}^{\mathrm{CW}}(\chi), \\
\mathcal{\tilde{W}}_{\mathrm{CCW}}^{( \pm ) }( \chi ) &=&%
\frac{1}{\pi p_{\pm }( t) }\sum\limits_{l,m,j,k=0}^{\infty } (-1)^{l}\Xi
_{m,k,m,j}^{( \pm ) }(t)\notag \\ && \times \mathcal{L}_{k,l}^{\mathrm{CCW}\dagger}(\chi)\mathcal{L}^{\mathrm{CCW}}_{j,l}(\chi),
\end{eqnarray}
\end{subequations}
where the terms $\mathcal{L}^{\epsilon=\mathrm{CW},\mathrm{CCW}}_{n,l}(\chi)$ can be calculated with Eq.~(\ref{663}).

\begin{figure*}[tbp]
\centering \includegraphics[width=0.98\textwidth]{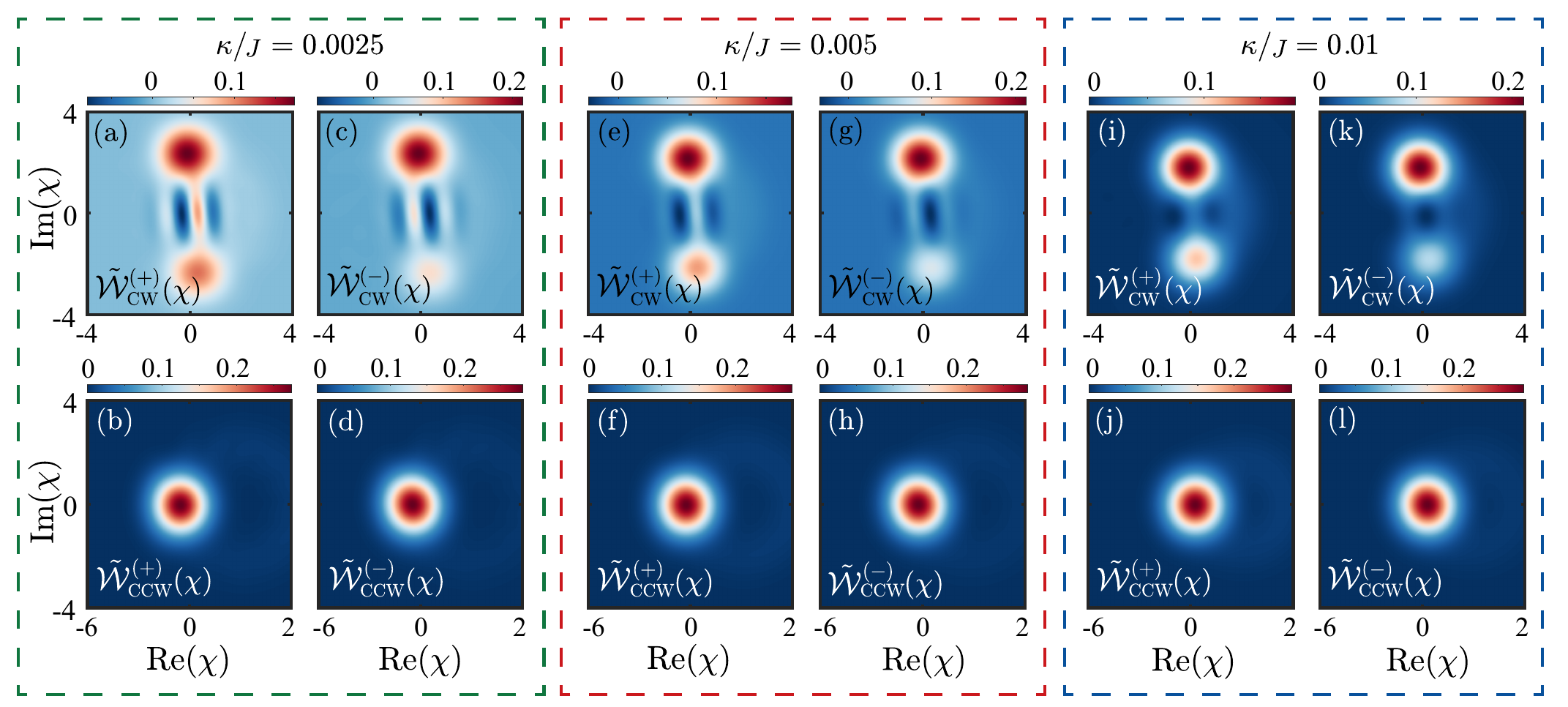}
\caption{Wigner functions $\mathcal{\tilde{W}}^{(\pm)}_{\mathrm{CW,CCW}}(\chi)$ corresponding to the states $\rho^{(\pm)}_{\epsilon=\mathrm{CW},\mathrm{CCW}}=\mathrm{Tr}_{\bar{\epsilon}}(\rho^{(\pm)})$, where $\rho^{(\pm)}$ are the states defined in Eq.~(\ref{eq33}). These Wigner functions are shown at the time $t_{s}=\pi/(2\zeta_{\mathrm{CW}})$ for various cavity-field decay rates: (a)$-$(d) $\kappa/J=0.0025$, (e)$-$(h) $\kappa/J=0.005$, and (i)$-$(l) $\kappa/J=0.01$. Other parameters used are the same as those in Fig.~\ref{Fig6}.}
\label{Fig8}
\end{figure*}
In Fig.~\ref{Fig8}, we show the Wigner functions $\mathcal{\tilde{W}}^{(\pm)}_{\mathrm{CW,CCW}}(\chi)$ of the generated states $\rho^{(\pm)}_{\epsilon=\mathrm{CW},\mathrm{CCW}}$ corresponding to different cavity-field decay rates. Compared to the closed-system results in Figs.~\ref{Fig5}(a) and~\ref{Fig5}(b), we observe that increasing the cavity-field decay rate reduces quantum coherence. For both the CW and CCW modes, when the decay rate takes $\kappa/J=0.0025$ [Figs.~\ref{Fig8}(a)$-$\ref{Fig8}(d)], $\kappa/J=0.005$ [Figs.~\ref{Fig8}(e)$-$\ref{Fig8}(h)], and  $\kappa/J=0.01$ [Figs.~\ref{Fig8}(i)$-$\ref{Fig8}(l)], we find that the interference fringes in the corresponding Wigner functions become weaker for a larger $\kappa$. Despite the reduction of quantum coherence due to dissipation, the CW mode remains in a cat state, while the CCW mode remains in a coherent state. This confirms that the chiral-cat-state generation scheme is robust even in the open-system cases.

\section{Discussions and conclusion \label{sec5}}
To analyze the experimental feasibility of our scheme, we present some  discussions concerning the experimental setups and the relevant parameters. Provided the use of spinning resonator, we focus on the microring resonator system coupled to a two-level system~\cite{Kavokin2007}. Currently, the strong coupling between a single atom and a monolithic microresonator has been observed~\cite{DayanSci2008,Aokiat2006}, with an effective coupling strength of $J/2\pi=50\pm12~\mathrm{MHz}$~\cite{Aokiat2006}. Moreover, the experimentally feasible parameters for the microresonator are $Q=6.0\times10^{9}$, $R=1.1$~mm, $n_{r}=1.4$, $\lambda=1550$~nm, and $\Omega=6.6$~kHz~\cite{MaayaniNa2018}. These parameters result in a cavity-field decay rate $\kappa\approx0.2$~MHz. The ratio of the cavity-field decay rate to the coupling strength used is of the same order as the experimental parameters. In our numerical simulations, the Sagnac-Fizeau frequency shift used is $\Delta_{\mathrm{sag}}=11J$, which corresponds to an angular velocity of $\Omega\approx\Delta_{\mathrm{sag}}c/[n_{r}R\omega_{c}(1-n_{r}^{-2})]=\Delta_{\mathrm{sag}}\lambda/[2\pi n_{r}R(1-n_{r}^{-2})]\approx2\pi\times136.9$~kHz for the above experimental parameters. We point out that this rotation angular velocity can be decreased by choosing a smaller coupling strength $J$. Recently, an angular velocity of $6.6$~kHz has been reported.  These analyses indicate that the suggested parameters should be accessible with the near future experimental technique. We also note that, even at lower angular velocities, we can still generate chiral cat states, where one mode remains in a coherent state, while the other mode is in a less distinguishable cat state.

In conclusion, we have proposed a reliable method for creating chiral cat states in a cavity-QED system consisting of a spinning resonator coupled to a two-level atom in the dispersive-coupling regime. We have shown that a cat state in the CW mode and a coherent state in the CCW mode can be successfully generated. By calculating the Wigner function, we have characterized quantum coherence properties of these generated cat states. In addition, we have examined the influence of the system dissipations on chiral-cat-state generation. Our work will contribute to the development of chiral quantum optics and the exploration of chiral optical devices.

\begin{acknowledgments}
J.-Q.L. was supported in part by the National Natural Science Foundation of China (Grants No.~12175061, No.~12247105, No.~11935006, and No.~12421005), the National Key Research and Development Program of China (Grant No.~2024YFE0102400), and the Hunan Provincial Major Sci-Tech Program (Grant No.~2023ZJ1010). H.J. is supported by the Natural Science Foundation of China (NSFC) (Grant No.~11935006), the Science and Technology Innovation Program of Hunan Province (Grant No.~2020RC4047), the National Key R\&D Program of China
 (Grant No.~2024YFE0102400), and the Hunan provincial major Sci-Tech program (Grant No.~2023ZJ1010). L.-M.K. is supported by the Natural Science Foundation of China (NSFC) (Grants No.~12247105, No.~12175060, No.~11935006, No.~12421005), the Hunan Provincial Major Sci-Tech Program (Grant No.~2023ZJ1010), and the XJ-Lab key project (Grant No.~23XJ02001). Y.-H.L. was supported in part by the Hunan Provincial Postgraduate Research and Innovation project (Grant No.~CX20240530).
\end{acknowledgments}

\end{document}